\documentclass[english,reprint]{revtex4-2}
\usepackage[T1]{fontenc}
\usepackage[latin9]{inputenc}
\usepackage{babel}
\usepackage{amsmath}
\usepackage{amssymb}
\usepackage{graphicx}
\usepackage[unicode=true]{hyperref}

\makeatletter
\hypersetup{colorlinks}

\makeatother

\begin{document}
\title{Collision of two self-trapped atomic matter wave packets in an optical
ring cavity}
\author{Jieli Qin}
\email{104531@gzhu.edu.cn; qinjieli@126.com}

\address{School of Physics and Materials Science, Guangzhou University, 230
Wai Huan Xi Road, Guangzhou Higher Education Mega Center, Guangzhou
510006, China}
\author{Lu Zhou}
\email{lzhou@phy.ecnu.edu.cn}

\address{Department of Physics, School of Physics and Electronic Science, East
China Normal University, Shanghai 200241, China}
\address{Collaborative Innovation Center of Extreme Optics, Shanxi University,
Taiyuan, Shanxi 030006, China}
\begin{abstract}
The interaction between atomic Bose-Einstein condensate (BEC) and light field in an optical ring cavity gives rise to many interesting phenomena such as supersolid and movable self-trapped matter wave packets. Here we examined the collision of two self-trapped atomic matter wave packets in an optical ring cavity, and abundant colliding phenomena have been found in the system. 
Depending on the magnitude of colliding velocity, the collision dynamics exhibit very different features compared with the cavity-free case.
When the initial colliding velocities of the two wave packets are small, they correlatedly oscillate around their initial equilibrium positions with a small amplitude. Increasing the collision velocity leads to severe scattering of the BEC atoms; after the collision, the two self-trapped wave packets usually break into small pieces. Interestingly, we found that such a medium velocity collision is of great phase sensitivity, which may make the system useful in precision matter wave interferometry. When the colliding velocity is further increased, in the bad cavity limit, the two wave packets collide phenomenally similar to two classical particles --- they firstly approach each other, then separate with their shape virtually maintained. However, beyond the bad cavity limit, they experience severe spatial spreading.
\end{abstract}
\maketitle

\section{Introduction}
For a cold atomic cloud and light field coupling system, the light field will put a mechanical potential on the atoms, and at the same time, the atomic cloud which plays the role of a medium also has a backaction on the light field. Therefore, the dynamics of the system usually show complex nonlinear features \citep{Ritsch2013Cold}, and abundant interesting phenomena can take place, such as soliton \citep{Dong2013Polaritonic,Qin2015Hybrid}, self-organization \citep{Larson2010Ultracold,Niedenzu2011Kinetic,Ostermann2015Atomic,Robb2015Quantum,Ostermann2016Spontaneous,Ostermann2017Probing}, supersolid \citep{Mivehvar2018Driven,Gietka2019Supersolid,Schuster2020Supersolid}, and many others \citep{Zhou2009Cavity,Zhou2010Spin,Zhou2011Cavity,Zhu2011Strong,Mendonca2012Photon,Diver2014Nonlinear,Norcia2018Cavity,Zhang2018Long,Landini2018Formation,Davis2019Photon,Guan2019Two,Ostermann2019Cavity}. Especially, in a ring cavity and Bose-Einstein condensate (BEC) coupling system, self-trapped wave packets can exist, and their moving dynamics have been examined recently \citep{Qin2020Self}. For the self-trapped wave packets, the collision dynamic is another problem of fundamental significance \citep{Gordon1983Interaction,Drazin1989Solitons,Aossey1992Properties,Papacharalampous2003Soliton,Rohrmann2013Two,Nguyen2014Collisions,Ferioli2019Collisions,Elhadj2020Singular}, and at the same time, their collisions may also found various applications, for example, logic gates implementation \citep{Jakubowski1997Information,Steiglitz2000Time,Kolokoltsev2002All,Kanna2003Exact,Rand2009Computing,Vijayajayanthi2018Harnessing}, entanglement generation \citep{Lewenstein2009Entanglement,Lai2009Entangled,Mishmash2009Quantum,Gertjerenken2013Generating} and soliton interferometry \citep{Polo2013Soliton,McDonald2014Bright,Helm2015Sagnac,Sakaguchi2016Matter,Wales2020Splitting}.

\begin{figure}
	\begin{centering}
		\includegraphics{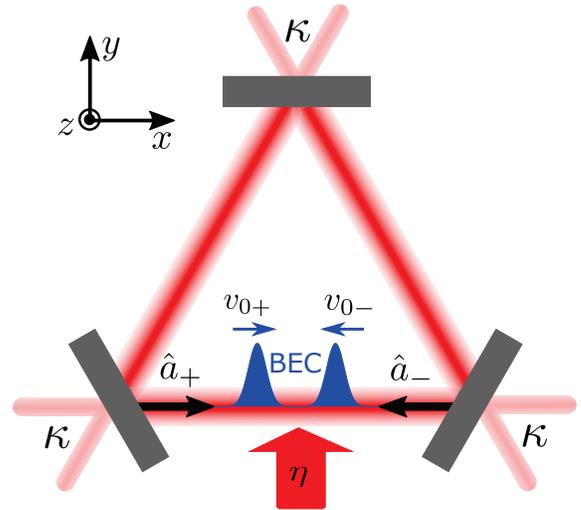}
	\end{centering}
	\caption{Diagram of the considered system. The ring cavity supports two degenerate 		counter-propagating modes ($\hat{a}_{+}$ and $\hat{a}_{-}$), and has a loss rate $\kappa$. A laser beam is transversely shined on the BEC atoms to pump the system, and the pumping strength is $\eta$. Two self-trapped atomic BEC wave packets with initial opposite velocities $v_{0 \pm}$ collide in the cavity, interacting with the light field.}
	\label{fig:Diagram}
\end{figure}

In this work, we study the collision of two self-trapped atomic matter wave packets in a ring-cavity, as shown in figure \ref{fig:Diagram}. A quasi-one-dimensional two-level atomic BEC is loaded into a ring cavity. The system is driven by transversely illuminating an off-resonant laser light on the BEC atoms (with detuning $\Delta_{a}$ and Rabi frequency $\Omega_{0}$), such that light fields of the two degenerate counter-propagating ring cavity modes ($\hat{a}_{\pm}e^{ik_{c}x}$ with $k_{c}$ being the wavenumber) are built up. The light field couples the transition between the two atomic energy levels (with strength $\mathcal{G}_{0}$), and forms an optical lattice potential for the BEC atoms. This optical lattice will feedbackly affect the dynamics of the condensate, and can support self-trapped matter wave packets \citep{Qin2020Self}. Moreover, because the optical lattice potential is produced by the BEC, the self-trapped BEC wave packet can move with this optical lattice being taken along \citep{Qin2020Self}. This gives us a good opportunity to examine the collision of two such self-trapped wave packets. 

We found that the collision of two such self-trapped wave packets shows rich phenomena under different colliding velocities. For a small velocity collision, the two wave packets only can oscillate around their initial equilibrium positions with a small amplitude. But one should note that due to the long range feature of the cavity light mediated interaction, the oscillations of the two wave packets are related rather than independent, thus the system may be used in the field of synchronizing matter wave oscillators. If the colliding velocity is increased, the BEC atoms are severely scattered, and after the collision, the two colliding wave packets usually break into many small pieces. However, under appropriate parameters, a considerable fraction of the atoms can be re-trapped. It is also found that this re-trapping phenomenon is quite sensitive to the initial relative phase between the two wave packets, indicating a potential application in realizing high precision matter wave interferometers. If the colliding velocity is further increased, the two self-trapped wave packets collide similar to two free classical particles --- they move close to one another, then separate again, and during the collision, their shape can be virtually maintained.

The rest of the paper is organized as follows: In section \ref{sec:Model}, the physical model is described, and the main formulae are presented. As the base to study the collision dynamics, the mobility of a single self-trapped matter wave packet is  briefly reviewed in section \ref{sec:Mobility}. Section \ref{sec:Collision} shows the main results of this work. In this section, the collision dynamics are studied both within (subsection \ref{subsec:BadCavityLimit}) and beyond (subsection \ref{subsec:BeyondBadCavityLimit}) the bad cavity limit. At last, the paper is summarized in section \ref{sec:Summary}.

\section{Model \label{sec:Model}}
The Hamiltonian of the considered system reads
\begin{equation}
\hat{H}=-\hbar\Delta_{c}\left(\hat{a}_{+}^{\dagger}\hat{a}_{+}+\hat{a}_{-}^{\dagger}\hat{a}_{-}\right)+\int\hat{\psi}^{\dagger}\hat{H}_{a}\hat{\psi}dx.\label{eq:H}
\end{equation}
The first term accounts for the light field of the two counter-propagating cavity modes, $\hat{a}_{+}$ ($\hat{a}_{+}^{\dagger}$) and $\hat{a}_{-}$ ($\hat{a}_{-}^{\dagger}$) are their annihilation (creation) operators, $\Delta_{c}$ is the detuning between the cavity modes and the pump laser, and $\hbar$ is the reduced Planck constant. While the second term is for the BEC and its interaction with the light field, where $\hat{\psi}$ is the atomic field operator, and $H_{a}$ is the single-atom Hamiltonian
\begin{equation}
\hat{H}_{a}=\frac{\hat{p}_{x}^{2}}{2m}+\hat{V}_{\mathrm{ac}}+\hat{V}_{\mathrm{ap}},\label{eq:Ha}
\end{equation}
with
\begin{equation}
\hat{V}_{\mathrm{ac}}=\hbar U_{0}\left[\hat{a}_{+}^{\dagger}\hat{a}_{+}+\hat{a}_{-}^{\dagger}\hat{a}_{-}+\left(\hat{a}_{+}^{\dagger}\hat{a}_{-}e^{-2ik_{c}x}+\mathrm{H.c.}\right)\right],\label{eq:Hac}
\end{equation}
\begin{equation}
\hat{V}_{\mathrm{ap}}=\hbar\eta_{0}\left[\hat{a}_{+}e^{ik_{c}x}+\hat{a}_{-}e^{-ik_{c}x}+\mathrm{H.c.}\right].\label{eq:Hap}
\end{equation}
Here, the physical meanings of the different terms are as follows: $\hat{p}_{x}^{2}/\left(2m\right)$ is the kinetic energy of a BEC atom with $\hat{p}_{x}=-i\hbar\partial_{x}$ being the momentum operator along the $x$-axis and $m$ being the mass of the atom. $\hat{V}_{\mathrm{ac}}$ is the optical potential caused by the two-photon scattering process between the two cavity modes, and its strength is $U_{0}=\hbar\mathcal{G}_{0}^{2}/\Delta_{a}$. $\hat{V}_{\mathrm{ap}}$is the optical potential caused by the two-photon scattering process between the pump light and one of the cavity modes, and $\eta_{0}=\hbar\mathcal{G}_{0}\Omega_{0}/\Delta_{a}$ is its strength, or in other words, the effective cavity pumping strength. For simplicity, we adopt natural units $m=\hbar=k_{c}=1$ in the following contents.

Applying the mean field theory, the quantum mechanical operators are approximately replaced by their $c$-number mean values (and here we also scale them by the total atom number $N$), i.e., $\hat{a}_{\pm}\rightarrow\alpha_{\pm}/\sqrt{N}$ and $\hat{\psi}\rightarrow\psi/\sqrt{N}$. Taking the mean values of the corresponding Heisenberg equations, the mean value variables obey equations
\begin{equation}
i\frac{\partial}{\partial t}\alpha_{\pm}=\left(-\Delta_{c}+U-i\kappa\right)\alpha_{\pm}+UN_{\pm2}\alpha_{\mp}+\eta N_{\pm1},\label{eq:alphapm}
\end{equation}
\begin{equation}
i\frac{\partial}{\partial t}\psi=\left(-\frac{1}{2}\frac{\partial^{2}}{\partial x^{2}}+V_{\mathrm{eff}}\right)\psi,\label{eq:NLSE}
\end{equation}
Here, we have included the cavity loss with decay rate $\kappa$ phenomenological. And for the neatness of the equations we also introduced some new
variables: $\eta=\sqrt{N}\eta_{0}$, $U=NU_{0}$, $N_{\pm1}=\int\left|\psi\right|^{2}\exp\left(\mp ix\right)dx$,
$N_{\pm2}=\int\left|\psi\right|^{2}\exp\left(\pm2ix\right)dx$, $V_{\mathrm{eff}}=V_{\mathrm{ac}}+V_{\mathrm{ap}}$,
$V_{\mathrm{ac}}=U\left(\left|\alpha_{+}\right|^{2}+\left|\alpha_{-}\right|^{2}\right)+U\left[\alpha_{+}^{*}\alpha_{-}\exp\left(-2ix\right)+\mathrm{c.c.}\right]$,
and $V_{\mathrm{ap}}=\eta\left[\alpha_{+}\exp\left(ix\right)+\alpha_{-}\exp\left(-ix\right)+\mathrm{c.c.}\right]$.

The steady state can be obtained by letting $\partial_t \alpha_{\pm} = 0$ and $\psi\left(x,t\right) = \psi\left(x\right) \exp\left(-i \mu t\right)$ with $\mu$ being the BEC chemical potential, that is
\begin{equation}
	\mu \psi\left(x\right) = \left[-\frac{1}{2}\frac{\partial^2}{\partial x^2} + V_\mathrm{eff} \left(x\right)\right] \psi\left(x\right),\label{eq:psiSteady}
\end{equation} 
\begin{equation}
	\alpha_{+}=-\frac{\left(-\Delta_{c}+U-i\kappa\right)\eta N_{+1}-\eta UN_{+2}N_{-1}}{\left(-\Delta_{c}+U-i\kappa\right)^{2}-U^{2}N_{-2}N_{+2}},\label{eq:alphap}
\end{equation}
\begin{equation}
	\alpha_{-}=-\frac{\left(-\Delta_{c}+U-i\kappa\right)\eta N_{-1}-\eta UN_{-2}N_{+1}}{\left(-\Delta_{c}+U-i\kappa\right)^{2}-U^{2}N_{2}N_{-2}}.\label{eq:alpham}
\end{equation}
From these equations, one sees that by transversely pumping the BEC an effective optical lattice potential can be built up. This effective potential can support a self-trapped BEC wave packet, and because this potential comes from the BEC, such a self-trapped wave packet is movable with the potential being taken along \citep{Qin2020Self}. In this work, we devote our efforts to the collision dynamics of two such wave packets.

Here, we also note that besides being used for calculating the steady state, equations  (\ref{eq:alphap}, \ref{eq:alpham}) are also used to describe dynamics of the light field in the bad cavity limit \citep{Cirac1995Laser,Horak2000Coherent}, which means that the cavity decay rate is much larger than the atom-cavity coupling, hence the cavity light field can quickly decay to the steady state. Mathematically this is to say that $\partial_t \alpha_{\pm} \approx 0$, thus equations (\ref{eq:alphap}, \ref{eq:alpham}) approximately holds, and they together with equation (\ref{eq:NLSE}) can describe the bad cavity limit dynamics [while beyond the bad cavity limit, the dynamics should be described by the original equations (\ref{eq:alphapm}, \ref{eq:NLSE})].

\begin{figure}
	\begin{centering}
		\includegraphics{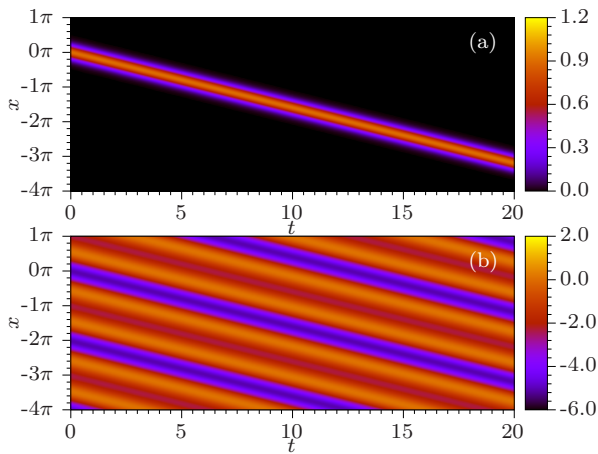}
	\end{centering}
	\caption{Mobility of a single self-trapped wave packet under the bad cavity limit. Top panel (a): Time evolution of the atomic density $\left|\psi\left(x,t\right)\right|^2$. Bottom panel (b): Time evolution of the corresponding effective optical lattice potential $V_\mathrm{eff}\left(x,t\right)$. Parameters used are $\Delta_c = -1$, $U=-0.5$, $\kappa=50$, $\eta=75$, and the initial velocity is a small one, $v_0 = 0.5$.}
\label{fig:MobilityBadCavityLimit}
\end{figure}
\begin{figure}
	\begin{centering}
		\includegraphics{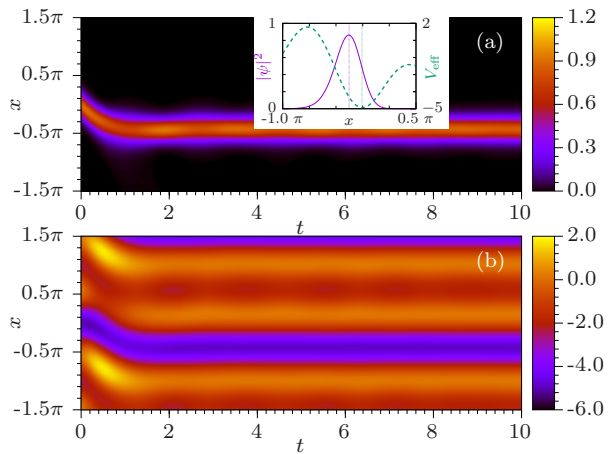}
	\end{centering}
	\caption{Mobility of a single self-trapped wave packet beyond the bad cavity limit. Top panel (a): $\left|\psi\left(x,t\right)\right|^2$. Bottom panel (b): $V_\mathrm{eff}\left(x,t\right)$. The inset shows the atomic density and effective potential at $t=0.4$, with the two vertical dotted lines indicating the center of the wave packet and bottom of the effective potential.  Parameters used are $\Delta_c = -1$, $U=-0.5$, $\kappa=5$, $\eta=7.5$, and the initial velocity is $v_0 = 2.5$.}
\label{fig:MobilityBeyondBadCavityLimit}
\end{figure}

\section{Mobility of a Single Self-Trapped Matter Wave Packet \label{sec:Mobility}}

Before dealing with the collision of two self-trapped matter-wave packets, let us first have a brief review on the results of mobility of a single self-trapped matter wave packet. Usually, the mobility of a wave packet in  a static optical lattice is determined by the Peierls-Nabarro barrier \citep{Dabrowska2004Interaction} --- when the velocity of a wave packet is higher than a critical value, it can move across the lattices; otherwise, it can not. However, for the now considering system, things are different. In the present system, the optical lattice is created by the BEC, when the BEC wave packet moves, the potential will also move accordingly. Especially, in the bad cavity limit, the lattice can instantaneously follow the movement of the BEC wave packet, thus the movement of the BEC wave packet will not be prevented no matter how small its velocity is, see figure \ref{fig:MobilityBadCavityLimit} where even though the velocity is set to a small value $v_0=0.5$, the wave packet moves freely.
 
Beyond the bad cavity limit, an initial moving wave packet undergoes a decelerating motion, see figure \ref{fig:MobilityBeyondBadCavityLimit}. In this case, the cavity light field can not instantaneously follow the dynamics of BEC, but some time is needed. When the BEC wave packet moves, the effective optical lattice can no longer follow exactly, instead, it falls some distance behind, as shown by the inset in panel (a). This falling behind potential put a friction force on the moving BEC wave packet, therefore it decelerates, and after traveling some distance it stops at a new steady-state location.

\section{Collision of Two Self-Trapped Matter Wave Packets \label{sec:Collision}}
In the last section, we show that in the ring cavity a single self-trapped matter wave packet is movable. This gives us a good opportunity to study the collision of two such self-trapped matter wave packets. We firstly find the self-trapped steady state $\psi_{0}$ (which consists of two localized wave packets centered at $-x_{0}$ and $+x_{0}$) using the imaginary time evolution method. Then, opposite initial velocities are given on the two wave packets by imprinting a positive velocity phase factor on the left side wave packet, while a negative velocity phase factor on the right side wave packet, i.e.,
\begin{equation}
\psi\left(x,t=0\right)=\psi_{0}\left(x\right) \exp\left[i\phi\left(x\right)\right],
\label{eq:InitialWave}
\end{equation}
with
\begin{equation}
\phi\left(x\right) = \left\{ 
\begin{matrix}
 v_{0+} x & + \ \Delta \phi&, & \qquad & x\leq0,\\
 v_{0-} x &             &, & \qquad & x>0.
\end{matrix}\right.
\label{eq:ImprintedPhase}
\end{equation}
Here $v_{0\pm}$ are the initial velocities of the two wave packets, and $\Delta \phi$ is an initial overall phase difference between the two wave packets. Since within or beyond the bad cavity limit the self-trapped wave packet moves in very different ways, next we examine the collision dynamics in these two cases separately. Within the bad cavity limit, the dynamics are simulated using equations (\ref{eq:NLSE}, \ref{eq:alphap}, \ref{eq:alpham}); while beyond the bad cavity limit, it is equations (\ref{eq:alphapm}, \ref{eq:NLSE}) been used.

\subsection{Bad Cavity Limit \label{subsec:BadCavityLimit}}
\subsubsection{Small Velocity Collision}
\begin{figure}
	\begin{centering}
		\includegraphics{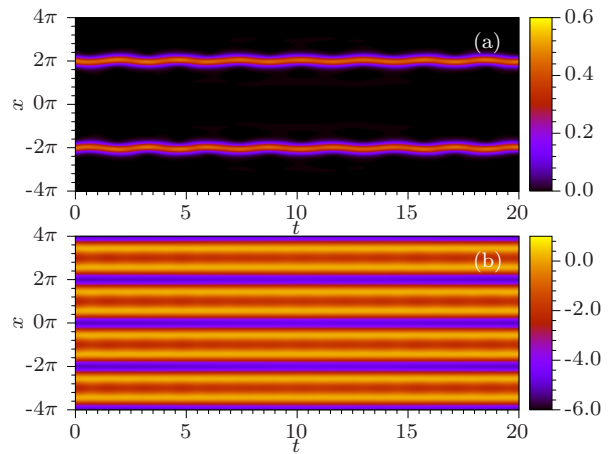}
	\end{centering}
	\caption{Small velocity collision of two self-trapped wave packets. Top panel (a): $\left|\psi\left(x,t\right)\right|^{2}$.Bottom panel (b): $V_{\mathrm{eff}}\left(x,t\right)$. Parameters used are $\Delta_{c}=-1$, $U=-0.5$, $\kappa=50$, $\eta=75$, $\Delta \phi=0$ and initial velocity is $v_{0 \pm}=\pm 0.5$.}
	\label{fig:SmallVelocity}
\end{figure}

Even though under the bad cavity limit a single self-trapped wave packet with a small initial velocity can move like a free particle [which has been shown in panels (a1,a2) of figure \ref{fig:MobilityBadCavityLimit}], two such wave packets will collide in a very different manner. In figure \ref{fig:SmallVelocity}, the collision of two self-trapped wave packets with initial velocity $v_{0 \pm}=\pm 0.5$ is numerically simulated. We see that this time the two wave packets can no longer move freely, but only oscillate around their initial equilibrium positions ($x=\pm 2\pi$) with a small amplitude [top panel (a)]. And looking for the effective optical potential, we found that it also no longer moves; instead, it becomes an almost static one [bottom panel (b)]. 

This phenomenon can be understood as follows. For simplicity, one imagines that the left direction moving wave packet $\Phi_{L}\left(x+x_{L,t}\right)$ ($x_{L,t}$ is the center of the wave packet at time $t$) creates a co-moving optical lattice potential $V_{L}\left(x,t\right)=-V_{0}\cos\left(x+x_{L,t}\right)$ (actually, the optical lattice has a more complex form, see the form of $V_{\mathrm{eff}}$, but we found that such a simple form already can qualitatively explain the phenomena we have observed). Similarly, the right direction moving wave packet  $\Phi_{R}\left(x+x_{R,t}\right)$ also creates a co-moving optical lattice potential $V_{R}\left(x,t\right)=-V_{0}\cos\left(x+x_{R,t}\right)$.  Then, the total effective optical lattice potential is the sum of $V_{L}$ and $V_{R}$, $V_{L+R}=-2V_{0}\cos\left[x+\left(x_{L,t}+x_{R,t}\right)/2\right]\cdot \cos\left[\left(x_{L,t}-x_{R,t}\right)/2\right]$. In the now case, the two wave packets only oscillate around their initial positions with a small amplitude, $x_{L,t} = - x_{R,t} \approx -2 \pi$, thus the total effective optical potential is approximately $V_{L+R}=-2V_{0}\cos\left(x\right)$ which is a static one. And in such a static optical lattice, wave packets with small initial velocity can only oscillate around their initial equilibrium positions. 

\begin{figure}
	\begin{centering}
		\includegraphics{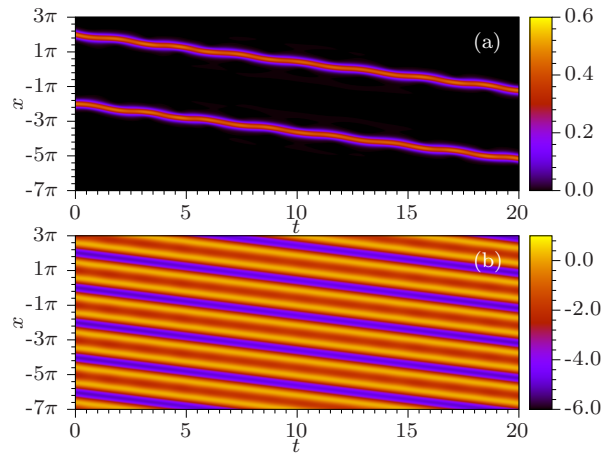}
	\end{centering}
	\caption{Collision of a small velocity moving self-trapped wave packet with an initial standstill one. Initially, a small velocity of $v_{0-}=-1.0$ is given to the wave packet at $x=2\pi$, while for the other one at $x=-2\pi$, no initial velocity is given, i.e., $v_{0+}=0$. }
	\label{fig:SmallVelocityDifferentVelocity}
\end{figure}

We emphasize that although the phenomenon presented here looks like the dipole oscillations of two wave packets in a static optical lattice, this is in fact an untouched collision phenomenon. In a true static optical lattice, two spatially separated wave packets undergo dipole oscillations independently. However, here the two wave packets contactless interact with each other via the cavity light field, such that their oscillations are not independent. For example, if initially we only give a small velocity to one of the wave packets, in a true static optical lattice only this wave packet will oscillate, the other one will keep still; but here this initially moving wave packet will drive the initial standstill wave packet to oscillate too, see figure \ref{fig:SmallVelocityDifferentVelocity}, where oscillations of both the two wave packets are obvious, and an overall free motion is also observed due to the initial velocity imbalance. The related motion of the two spatially separated wave packets in this case might find applications in synchronizing matter wave oscillators \citep{Samoylova2015Synchronization, Li2017Quantum}.

\begin{figure}
	\begin{centering}
		\includegraphics{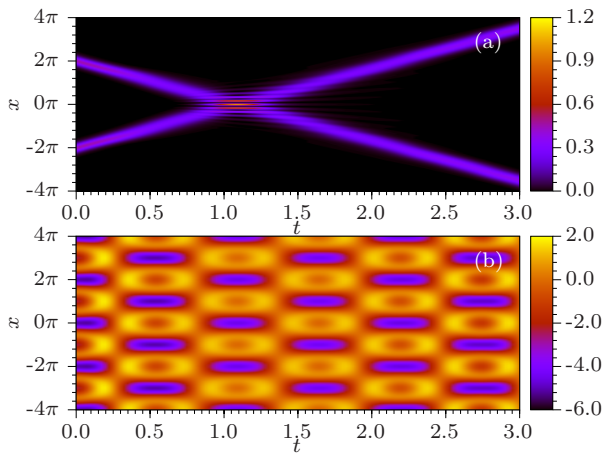}
	\end{centering}
	\caption{Large velocity collision of two self-trapped wave packets. The same as figure \ref{fig:SmallVelocity}, except that the initial velocities of the two wave packets are set to $v_{0 \pm}=\pm 6.0$.}
    \label{fig:LargeVelocity}
\end{figure}

\subsubsection{Large Velocity Collision}
Next, we deal with the large velocity collision. In figure \ref{fig:LargeVelocity}, we set the initial velocity to $v_{0\pm}=\pm 6$, so as to let the kinetic energy $v_{0\pm}^{2}/2=18$ be larger than the initial effective optical lattice depth ($V_{\mathrm{eff}}^{\mathrm{depth}}\approx 5$). We found that in such a case the two wave packets behave like two classical particles, they firstly approach each other, then they split again with their shape virtually maintained. And during the overlapping time, due to their wave natures an interference pattern is formed, see the top panel (a). This may remind one of the collision of two BEC bright solitons \citep{Nguyen2014Collisions}, which is phenomenally very similar. However, we note that for the bright solitons their shapes are exactly maintained during the evolution, but here the shape maintenance is only ``virtually'', if one examines the very detail of figure \ref{fig:LargeVelocity}, very slightly broadening of the wave packets can be observed. 

And, the physics behind is also quite different. In the case of BEC bright soliton, the shape maintaining collision is due to the integrability of the system. Here the phenomenon should be understood by examining the corresponding effective optical lattice potential. The simple analytical discussion in the previous small velocity collision case also adopts, the two wave packets still create a total effective optical lattice potential $V_{L+R}=-2V_{0}\cos\left[x+\left(x_{L,t}+x_{R,t}\right)/2\right] \cdot  \cos\left[\left(x_{L,t}-x_{R,t}\right)/2\right]$. But this time the centers of the wave packets move freely, that is $x_{L,t}=-2\pi + v_{0}t$ and $x_{R,t}=2\pi - v_{0} t$, therefore the total effective potential is simplified to $V_{L+R}=-2V_0 \cos\left(x\right) \cos\left(v_{0} t\right)$. This formula implies that every time $t$ passes through the value of $\left(m+1/2\right)\pi/v_0$  with $m$ being an integer number, the minimums of the total lattice potential change to maximums, and vice versa. This effectively causes the optical lattice potential to co-move with the two wave packets, as shown in the bottom panel (b). As a result, the two wave packets can virtually maintain their shapes during the evolution.

\begin{figure*}
    \begin{centering}
        \includegraphics{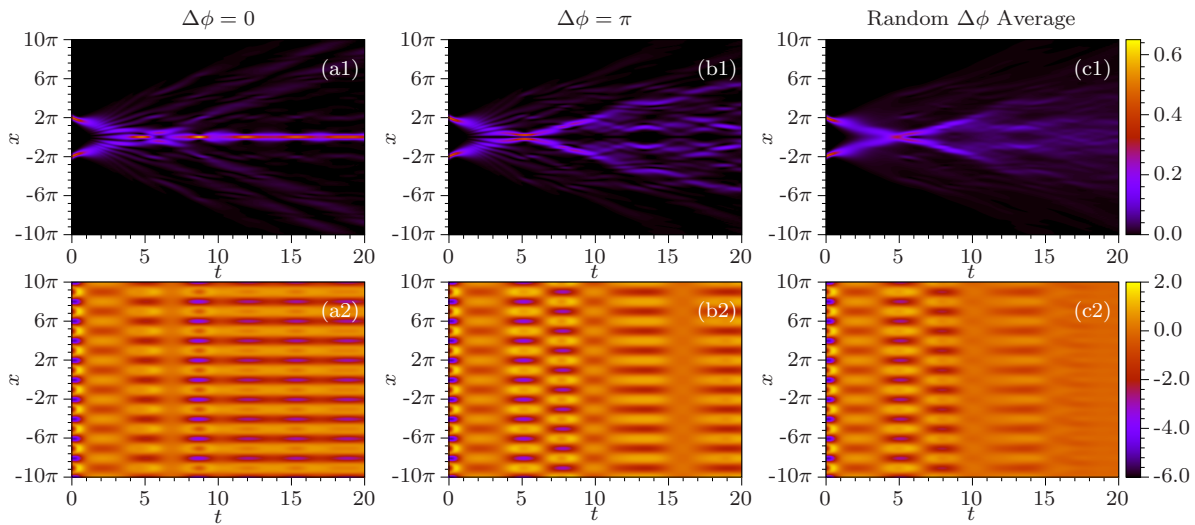}
    \end{centering}
    \caption{Medium velocity collision of two self-trapped wave packets. Top panels (a1-c1): $\left|\psi\left(x,t\right)\right|^2$. Bottom Panels (a2-c2): $V_{\mathrm{eff}}\left(x,t\right)$. Panels (a1,a2): In-phase ($\Delta \phi = 0$) collision dynamics. Panels (b1,b2): Opposite-phase ($\Delta \phi=\pi$) collision dynamics. Panels (c1,c2): Average of $\left|\psi\left(x,t\right)\right|^2$ and $V_{\mathrm{eff}}\left(x,t\right)$ over 256 simulations with random values of relative phase $\Delta \phi$. Other parameters used are $\Delta_c=-1$, $U=-0.5$, $\kappa=50$, $\eta=75$, $v_{0\pm}=\pm 2.5$.}
    \label{fig:MediumVelocity}
\end{figure*}

\begin{figure}
	\begin{centering}
		\includegraphics{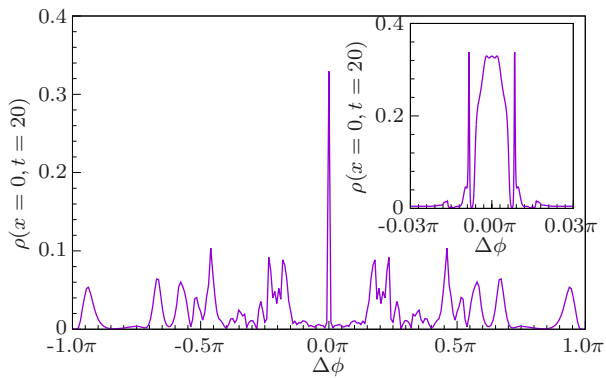}
	\end{centering}
	\caption{Sensitivity of the medium velocity collision with respect to the relative phase. Atomic density $\rho\left(x=0,t=20\right)$ as a function of $\Delta \phi$ is plotted. The inset is a detailed plot of the curve around $\Delta \phi=0$. Parameters used are $\Delta_c=-1$, $U=-0.5$, $\kappa=50$, $\eta=75$, $v_{0\pm}=\pm 2.5$.}
	\label{fig:Rho0VSPhase}
\end{figure}

\begin{figure}
	\begin{centering}
		\includegraphics{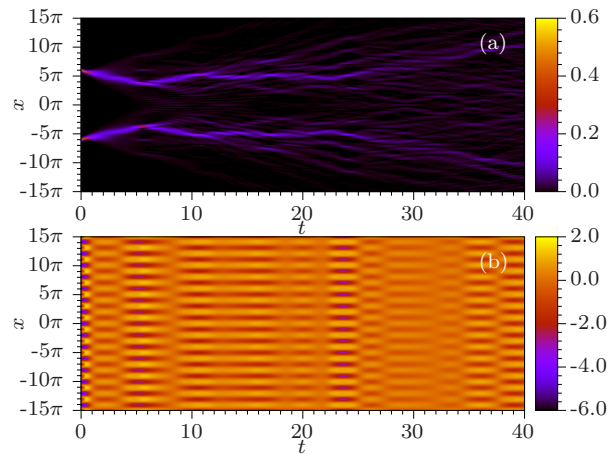}
	\end{centering}
	\caption{Medium velocity collision of two self-trapped wave packets with a large initial spatial separation. The same as panels (a1,a2) of figure \ref{fig:MediumVelocity} except that the initial spatial separation of the two wave packets is set to a value of $12\pi$ which is 3 times as large as that in figure \ref{fig:MediumVelocity}.}
	\label{fig:MediumVelocityLargeSeparation}
\end{figure}

\subsubsection{Medium Velocity Collision}
The medium velocity collision shows absolutely new features, as shown in figure \ref{fig:MediumVelocity}, where we set the collision velocity to be $v_{0\pm}=\pm 2.5$. For such a collision velocity, the kinetic energy ($v_{0\pm}^{2}/2=3.125$) is a little smaller than the depth of the initial effective optical lattice ($V_{\mathrm{eff}}^{\mathrm{depth}}\approx5$). But as the wave packets begin to move, the depth of effective optical lattice will be reduced. Therefore, they can overcome the lattice barrier and move close to each other. However, this time the two wave packets no longer can separate again with their shape maintained. In fact, the atoms are severely scattered, the self-tapped wave packets break into many small pieces and widely spread in the space.

The lack of wave packet shape maintenance in this case may be regarded as a drawback at the first sight. However, we found that this can prolong the time that interference can be observed. In the shape-maintaining large velocity collision, the two wave packets overlap for a very short time interval, when they separate again, the interference pattern disappears. But here, the interference can last for quite a long time. In panels (a1,b1), the evolution of atomic density $\left|\psi\left(x,t\right)\right|^2$ for relative phase $\Delta \phi = 0$ and $\Delta \phi = \pi$ are plotted. And for comparison in panel (c1), we show the average atomic density over 256 simulations with random values of the initial relative phase $\Delta \phi$. Since the random phase average will wipe out the interference pattern, the difference between panels (a1,b1) and (c1) indicates the effect of interference. Comparing these panels, we see that interference can be observed from the time of about $t=2.5$ to the end of the numerical simulation  $t=20$. 

One may also expect that a remote interference \citep{Saba2005Light} of the two wave packets can be observed even before they meet, due to the long-range feature of the cavity light field mediated interaction. But this does not happen, indicated by the fact that before the two wave packets meet ($t<2$), panels (a1,b1) are the same as panel (c1). It is because that the cavity light field is determined by the atomic density $\left|\psi\left(x,t\right)\right|^2$, having nothing to do with the phase of the BEC wave function, see the formulae in section \ref{sec:Model}.

The most interesting phenomenon in the medium velocity collision is the re-trapping of BEC atoms shown in panel (a1) ---  after the two wave packets collide, a considerable fraction of the atoms are re-trapped around $x=0$, forming a new long time stable wave packet. We attribute this phenomenon to the dynamical property of the optical lattice potential. Because when we simulate the collision of two matter wave packets in a static optical lattice, such a phenomenon is not observed. We also found that this phenomenon is extremely sensitive to the initial relative phase $\Delta \phi$ between the two wave packets. As shown in figure \ref{fig:Rho0VSPhase}, when $\Delta \phi$ slightly departs from the value of 0, the atomic density at $x=0, t=20$ sharply drops from its maximum to almost zero. This sensitivity of atomic density with respect to the relative phase between two initially spatial separated matter wave packets may make the system useful in realizing a high precision matter wave interferometer \citep{Polo2013Soliton,McDonald2014Bright,Helm2015Sagnac,Sakaguchi2016Matter,Wales2020Splitting}.

In this case, we also found that the initial spatial separation between the two wave packets also substantially affects the collision dynamics.  In figure \ref{fig:MediumVelocityLargeSeparation}, we keep other parameters exactly the same as in panels (a1,a2) of figure \ref{fig:MediumVelocity}, only triple the initial spatial separation between the two wave packets, and very different collision dynamics are observed --- strong scattering of the atoms occurs even before the two wave packets meet each other. This is because the two wave packets can interact indirectly via the cavity light field even though they are spatially separated, i.e., the cavity mediated interaction is a long range one \cite{Mottl2012Roton, Joshi2015Cavity, Cheng2021Bose}.

At the last of this subsection, we make two notes. Firstly, in the small/large collision case, we also studied the effect of initial spatial separation and relative phase $\Delta \phi$ between the two wave packets. We found that the initial spatial separation hardly affects the dynamics for both cases, and the relative phase only shifts the interference stripe during the overlapping time of the two wave packets in the large velocity case. No other interesting new phenomenon has been observed. Secondly, for the medium velocity collision, the matter wave packets and the effective optical lattice evolve in such a complex way, as a result, we fail to find a simple analytical explanation of the dynamics like that in the small/large velocity case.

\subsection{Beyond Bad Cavity Limit \label{subsec:BeyondBadCavityLimit}}
\begin{figure}
    \begin{centering}
        \includegraphics{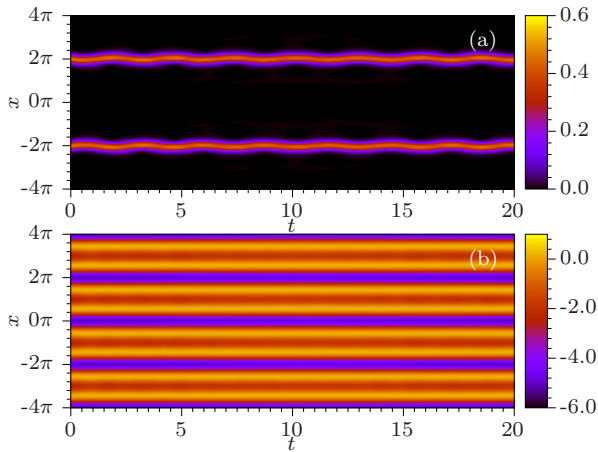}
    \end{centering}
    \caption{Small velocity collision of two self-trapped wave packets beyond the bad cavity limit. Similar to figure \ref{fig:SmallVelocity}, except that the simulation is done beyond the bad cavity limit with parameters $\Delta_{c}=-1$, $U=-0.5$, $\kappa=5$, $\eta=7.5$, $\Delta \phi=0$, and initial velocity $v_{0\pm}= \pm 0.5$.}
    \label{fig:SmallVelocityBeyondBadCavityLimit}
\end{figure}

\begin{figure}
    \begin{centering}
        \includegraphics{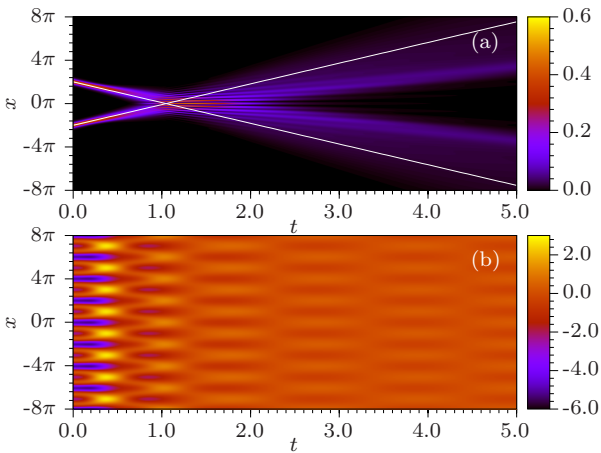}
    \end{centering}
    \caption{Large velocity collision of two self-trapped wave packets beyond the bad cavity limit. Similar to figure \ref{fig:LargeVelocity}, except that the simulation is done beyond the bad cavity limit with parameters $\Delta_{c}=-1$, $U=-0.5$, $\kappa=5$, $\eta=7.5$, $\Delta \phi=0$, and initial velocity $v_{0\pm}= \pm 6.0$. The two white lines show the free particle moving trajectories for reference.}
    \label{fig:LargeVelocityBeyondBadCavityLimit}
\end{figure}  

\subsubsection{Small Velocity Collision}
Recall that beyond the bad cavity limit, a single moving self-trapped wave packet feels a friction force from the optical potential, as a result its motion quickly stops. This may lead one to think that beyond the bad cavity limit, oscillation of the two wave packets during a small velocity collision will also quickly vanish. But in fact, we found that the oscillation is hardly damped, as shown in figure \ref{fig:SmallVelocityBeyondBadCavityLimit}. This is because the small amplitude oscillations of the two wave packets produce an almost static optical lattice (this has been explained in the case of bad cavity limit, and here it also holds), and a static potential will have no friction effect on the wave packets, thus the oscillation does not damp.

\subsubsection{Large Velocity Collision}
The large velocity collision dynamics beyond the bad cavity limit are shown in figure \ref{fig:LargeVelocityBeyondBadCavityLimit}. In the top panel (a), for reference the two white lines show the free particle moving trajectories, $x = \pm x_{0} - v_{0\pm}t = \pm 2\pi \mp 6t$. Comparing the evolution of the two wave packets with these two lines, the effect of friction obviously shows. Compared with the bad cavity limit figure \ref{fig:LargeVelocity}, another feature in this figure is the spatially spreading of the wave packets. Within the bad cavity limit, moving of the wave packets will reduce the depth of the effective lattice, but as the moving continues, the lattice quickly recovers, thus the wave packets hardly spread. But beyond the bad cavity limit, the optical lattice can no longer recover in time, therefore the wave packets can not be well trapped, and they spread. Even worse, as the wave packets spread, according to formulae in section \ref{sec:Model}, the values of $N_{\pm 1, \pm 2}$ become small, and the depth of the optical lattice will be further reduced. Thus, as time goes the potential becomes weaker and weaker, see bottom panel (b), and the wave packets spread wider and wider.

\subsubsection{Medium Velocity Collision}
\begin{figure*}
	\begin{centering}
		\includegraphics{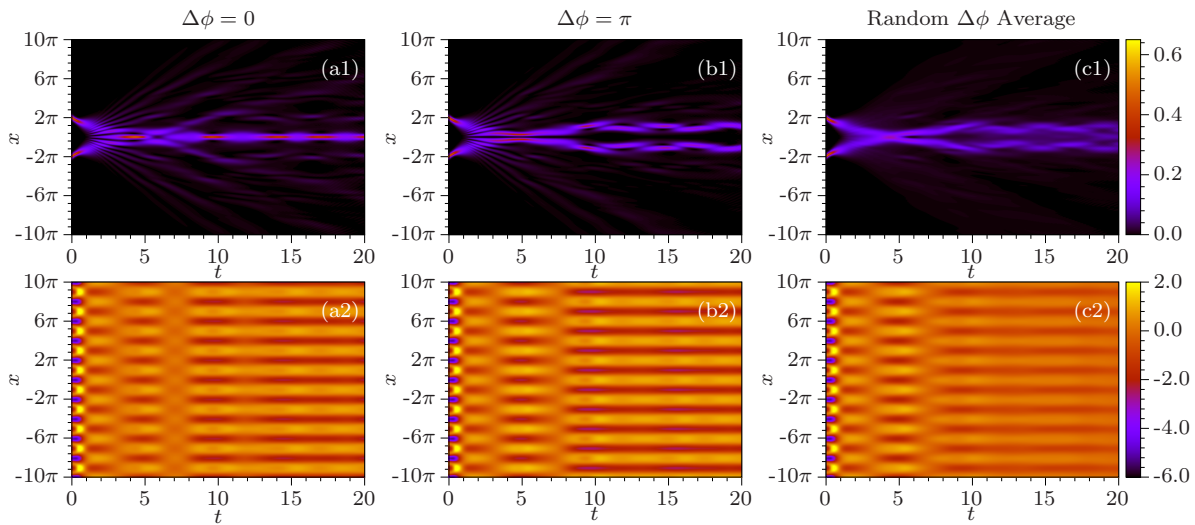}
	\end{centering}
	\caption{Medium velocity collision of two self-trapped wave packets beyond the bad cavity limit. Similar to figure \ref{fig:MediumVelocity} except that the simulation is done beyond bad cavity limit with parameters $\Delta_c=-1$, $U=-0.5$, $\kappa=5$, $\eta=7.5$, $v_{0\pm}=\pm 3.5$.}
	\label{fig:MediumVelocityBeyondBadCavityLimit}
\end{figure*}

\begin{figure}
	\begin{centering}
		\includegraphics{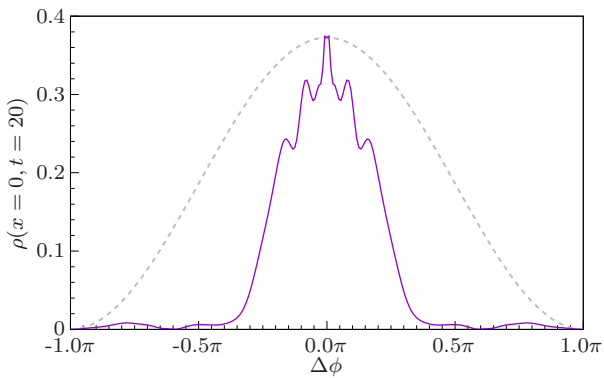}
	\end{centering}
	\caption{Sensitivity of the medium velocity collision with respect to relative phase beyond the bad cavity limit. Atomic density $\rho\left(x=0,t=20\right)$ vs $\Delta \phi_0$ is plotted as the solid violet line. Parameters used are $\Delta_c=-1$, $U=-0.5$, $\kappa=5$, $\eta=7.5$, $v_{0\pm}=\pm 3.5$. For comparison, the gray dashed line shows the phase sensitivity of an usual double slit interference, $\rho = \rho_0 \left[1+\cos\left(\Delta \phi\right)\right]/2$.}
	\label{fig:Rho0VSPhaseBeyondBadCavityLimit}
\end{figure}

Beyond the bad cavity limit, the re-trapping phenomenon of matter wave after a medium velocity collision can also be observed, as shown in panels (a1,a2) of figure  \ref{fig:MediumVelocityBeyondBadCavityLimit}. But due to the friction force, this time a larger initial velocity is needed to be given on the wave packets (here $v_0 = 3.5$, compared to $v_0 = 2.5$ in the bad cavity limit). And because of the friction force, this time for other phases $\Delta \phi \neq 0$, the atoms also can not move very far away, thus in panels (b1,c1) we also observe that a large fraction of the atoms distribute around the region near $x=0$. This will leads to a loss of the phase sensitivity. As shown in figure \ref{fig:Rho0VSPhaseBeyondBadCavityLimit}, this time the violet solid curve of $\rho\left(x=0,t=20\right)$ vs $\Delta \phi$ is much broader than that in the bad cavity limit case figure \ref{fig:Rho0VSPhase}. However, even though the phase sensitivity has been lost to a certain extent, it is still better than that of a usual double slits interference which is $\rho = \rho_0 \left[1+\cos\left(\Delta \phi\right)\right]/2$, shown as the gray dashed line in the same figure.


\section{Summary \label{sec:Summary}}
In summary, we found very rich colliding phenomena of two self-trapped BEC wave packets in a ring cavity, due to the complex interplay between the BEC and optical lattice potential induced by it. The colliding dynamics strongly depend on the collision velocity. In the small velocity collision case, the induced optical lattice is almost static, the wave packets do not have enough kinetic energy to overcome the lattice barrier, thus they only oscillate around their initial equilibrium position with a small amplitude. Because the cavity mediated interaction is long range, even though the two wave packets are contactless, their oscillations are related rather than independent, this indicates a possibility of realizing matter wave oscillator synchronization. In the medium velocity collision case, the wave packets and the induced optical lattice potential strongly affect each other, and this leads to a severe scattering of the BEC atoms. Under appropriate parameters, parts of the atoms can be re-trapped after the collision. Interestingly, we found that the re-trapped atomic density is very sensitive with respect to the initial phase difference of the two wave packets. This implies a high precision matter wave interferometer may be built using the present system. And in the large velocity collision case, the two wave packets can collide with their shape virtually maintained under the bad cavity limit; while beyond the bad cavity limit, the wave packets suffer severe spatial spreading.

Here, we consider the system within the mean field theory. It is known that quantum fluctuation of both the matter wave field \citep{Zin2006Elastic, Deuar2011Mean} and cavity light field \citep{Szirmai2009} will induce a depletion of the BEC atoms. And, the scattered atoms can possess interesting properties such as correlation \citep{Deuar2007Correlations, Krachmalnicoff2010Spontaneous}, entanglement \citep{Joshi2015Cavity} and squeezing \citep{Deuar2013Tradeoffs} in the collision process. This would be a promising future researching direction.

\begin{acknowledgments}
The authors acknowledge supports from the National Natural Science Foundation of China (Grants No. 11904063, No. 12074120, No. 11847059, and No. 11374003), and the Science and Technology Commission of Shanghai Municipality (Grant No. 20ZR1418500).
\end{acknowledgments}

\end{document}